\pdfoutput=1
\documentclass[aps,twocolumn,pra,tightenlines,floatfix,showpacs,superscriptaddress]{revtex4-1}
\usepackage{graphicx}
\usepackage{epstopdf}
\usepackage[english]{babel}
\usepackage{amsmath}
\usepackage{amssymb}
\usepackage{mathtools}
\usepackage{times}
\usepackage{appendix}
\usepackage{bm}
\usepackage{float}
\usepackage{color}
\usepackage{longtable}
\usepackage{url}
\usepackage[usenames,dvipsnames]{xcolor}
\usepackage[colorlinks=true,linkcolor=Blue,urlcolor=Blue,citecolor=Blue]{hyperref}

\graphicspath{{./Figures}}
\newcommand{\RNum}[1]{\uppercase\expandafter{\romannumeral #1\relax}}    

\begin{document}

\title{Multiple-parameter estimation in a sagnac interferometer}

\author{Xu Yu}
\affiliation{Zhejiang Institute of Modern Physics, Department of Physics, Zhejiang University, Hangzhou 310027, China}

\author{Hongbin Liang}
\affiliation{Zhejiang Institute of Modern Physics, Department of Physics, Zhejiang University, Hangzhou 310027, China}

\author{Xiaoguang Wang}
\email{xgwang1208@zju.edu.cn}
\affiliation{Zhejiang Institute of Modern Physics, Department of Physics, Zhejiang
University, Hangzhou 310027, China}

\date{\today}
\begin{abstract}
We explored the general characteristics of a Sagnac interferometer in a multiparameter estimation process. We find that in the two-parameter estimation scenario, one cannot make both parameter measurement results reach the Heisenberg limit (HL) simultaneously when the input resources are maximally entangled. Only one of the parameters' uncertainty can approach the HL while the other is only scaled by the standard quantum limit (SQL).
We also discussed the constraint conditions that make the quantum Cram${\rm\acute{e}}$r-Rao bound saturable. These constraint conditions would prompt one to choose proper evolution time and optimal input state.
Under the constraint conditions, we find that the HL result obtained in the two parameter scenario would catch up with or even be more precise than that acquired by the single parameter measurement process in some special cases. Such general features about the Sagnac system revealed in our work may have a reference value in actual experiments.
\end{abstract}

\maketitle

\section{Introduction}
\vspace*{-1.5ex}
Discussion about enhancing quantum estimation is continuously popular in the recent decades~\cite{Caves1981,Pezze2008,Zheng2015,Hradil1997,Holland1993,Strobel2014,Joo2011,Cooper2010,Hyllus2010,Yu2018,Boixo2007}. However, works in this field done before are mostly focused on individual parameter metrology \cite{Joo2011,Cooper2010,Hyllus2010,Yu2018,Boixo2007,Liu2013,Lee2002,Liu2015,Luo2017,Yao2019}. While in many actual estimation tasks, such as detecting weak gravitational field\cite{Troja2017,Abbott2017}, there would exist more than one unknown parameters that are correlated with each other, which requires one to measure them simultaneously. Hence, discussions about multiparameter estimation has become popular in recent years\cite{Yang2019,Liu2016,Liu2019,Szczykulska2016,Humphreys2013,Baumgratz2016}. Except for the need of actual experiments, there also exists many other advantages of multiparameter estimation in quantum metrology. For example, researches have shown that parameter estimation in multiparameter scenario would provide more precise result than that in single parameter scheme in some cases\cite{Humphreys2013,Baumgratz2016}. Additionally, the quantum Fisher information matrix (QFIM) that scales the limit a the multiparameter estimation result can also describe the distance between two quantum states\cite{Pires2016,Samuel1994,Taddei2013}, therefore, the results obtained by multiparameter estimation would also have a geometrical characteristic that can figure the evolution of a quantum system\cite{Pires2016,Samuel1994,Taddei2013}.
\par

Although have many advantages, there also exists some incompletely solved problems that constraint the application of the multiparameter estimation skill. Both of the single and multiple parameter estimations satisfy the Cram$\rm{\acute{e}}$r-Rao bound, but unlike the individual parameter estimation that its ultimate precision can always be obtained\cite{Braunstein1994}, the lower limit of the precision results of all parameters in the multiparameter scenario usually cannot be approached simultaneously\cite{Pezze2017,Hayashi2008,Ragy2016}, in other words, the Cram$\rm{\acute{e}}$r-Rao bound cannot be saturated unconditionally\cite{Matsumoto2002,Fujiwara2001}. Then how can all these precision limits be obtained synchronously is of great concern to researchers. Our work in this note also discussed this problem. We find the constraint conditions that saturate the Cram$\acute{e}$-Rao bound and calculate the ultimate measurement precision in a general input state scenario. We choose a Sagnac interferometer as the measurement apparatus.
\par
The main part of a Sagnac interferometer is a cyclic structure that can transport light or trapped particles. If two parts of light or particles counttransport along a rotating Sagnac apparatus, there would generate a relative phase which is called the Sagnac phase between the two parts when they recombined again, just as shown in Fig.1. Such an effect is very useful in quantum estimation experiments\cite{Gustavson1997,Starling2010,Arditty1981}, adding the fact that preparing a cyclic interferometer is not difficult under current technical conditions, recently many works about quantum estimation are done on the basis of the Sagnac interferometer\cite{Luo2017,Yao2019,Stevenson2015,Che2018,Kim2004,Riehle1991,Hasselbach1993}, so do us in this paper. We discussed the performance of a Sagnac interferometer in the multiparameter estimation process. We find the measurement precision is strongly limited by the interferometer's inherent features. Our results may be helpful to these that using Sagnac interferometer to measure parameters.
\par
This paper is organized as follows. In Sec.~(\ref{background}), we simply introduce the structure of a Sagnac interferometer and some theoretical backgrounds of multiparameter measurement. In Sec.~(\ref{calculating}), we calculate the generators of the parameters through an evolution operator, we furthermore derive the general expressions of the elements of the quantum Fisher information matrix, we also confirm the constraint conditions that makes both parameters get the ultimate measurement precision simultaneously. In Sec.~(\ref{general analysis}), we analyse the measurement results in an overall perspective. We find the two parameter uncertainties cannot reach the HL at the same time, only one can approach the HL while the other is only limited by the SQL. In Sec.~(\ref{CC1}) and ~(\ref{CC2}), we further analyse the measurement precision by considering the constraint conditions.

\begin{figure}[t]
\includegraphics[width=3.3in,clip]{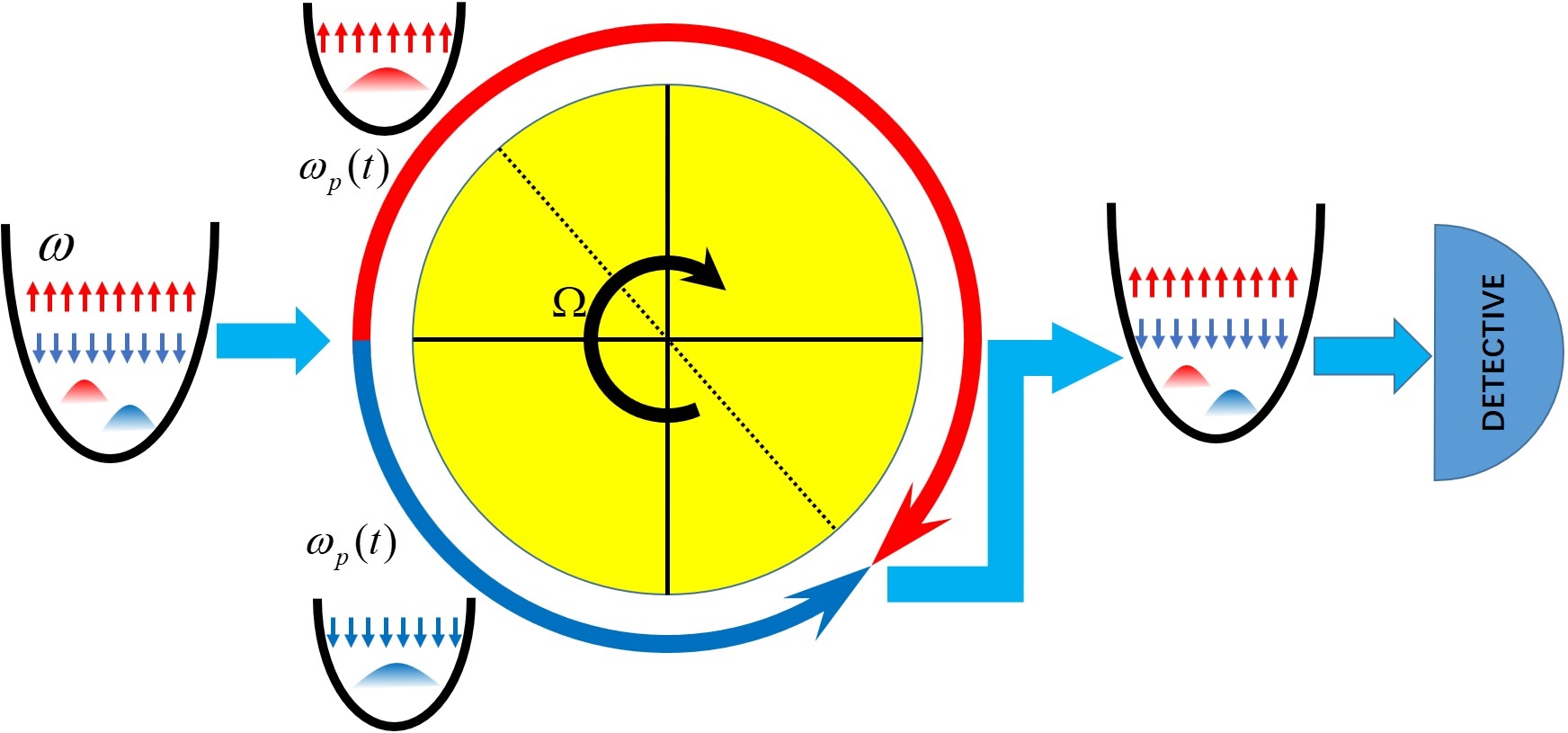}
\caption{Schematic diagram of the parameter measurement in a Sagnac interferometry. The Sagnac system here is simplified as a clockwise rotating disc with uniform angular velocity $\Omega$. The input state which is prepared by trapping N spin-1/2 particles in a harmonic potential with frequency $\omega$ is injected into the Sagnac system and transformed by driving the spin-up and spin-down particles countertransport along the disc with sweeping angular frequency $\omega_p(t)$ in the rotating frame. The output state containing the information of $\omega$ and $\Omega$ is formed when the particles recombine again. One can readout these parameters by detecting the output state. }
\label{Fig:1}
\end{figure}

\vspace*{-1.5ex}
\section{Sagnac Interferometry Model And Multiparameter Measurement Background}
\vspace*{-1.5ex}
\label{background}
We prepared the input state by trapping N spin-1/2 particles in a harmonic potential, each of them has half probability of spin up or down. Such a state is maximally entangled and can be generated by applying a resonant $\pi/2$ pulse to the N spin-down particles\cite{Stevenson2015}. We write the input state as~\cite{Luo2017,Yao2019}
\begin{eqnarray}
\label{input1}
|\psi_0\rangle=\frac{1}{\sqrt2}\left( \bigotimes\limits_{k=1}^{N}|\psi_{\uparrow}\rangle
_{k}|\uparrow\rangle_k+\bigotimes\limits_{k=1}^{N}|\psi_{\downarrow}\rangle
_{k}|\downarrow\rangle_k \right),
\end{eqnarray}
where $|\psi\rangle_k$ is the $k'$th particle's position state in the harmonic potential. In this paper, we assume the position states of the particles with uniform spin direction has the same form, i.e.,$|\psi_\uparrow\rangle_k=|\psi_\uparrow\rangle$, $|\psi_\downarrow\rangle_k=|\psi_\downarrow\rangle$. When the input particles encounter with the Sagnac system, the spin-up and down ones are driven to counter-transport along a circular path of radius $R$ with angular velocity $\omega_p(t)\pm\Omega$ in the laboratory frame, as seen in Fig.1. The effect of the Sagnac interferometer to the input resource is characterized by the Hamiltonian\cite{Luo2017}
\begin{align}
\label{H1}
H(t)=\sum\limits_{k=1}^{N}\hbar\omega a_k^\dagger a_k+i\hbar\mu\sqrt{\omega}\left(a_k-a_k^\dagger\right)\left(\Omega+\sigma_z^{(k)}\omega_p(t)\right),
\end{align}
where $\mu=\sqrt{m/(2\hbar)}R$, with $m$ is the mass of a single particle, $a_k^\dagger(a_k)$ is the creation (annihilation) operator of the $k$'th particle for the position state in the harmonic trap, $\sigma_z^{(k)}=|\uparrow\rangle_k\langle\uparrow|
-|\downarrow\rangle_k\langle\downarrow|$ is the pseudo-spin operator of the $k$'th particle, $\omega$ is the frequency of the harmonic potential, $\Omega$ is the angular velocity of the Sagnac system and $\omega_p(t)$ is the relative angular velocity between the trapped particles and the rotating Sagnac interferometer. $\omega$ and $\Omega$ are the two parameters we intend to estimate. Supposing $t=0$ is the time when particles enter the Sagnac system and then split into two parts and recombined again at time $\tau$. As the two opposite paths forms a complete circle, $\tau$ should satisfy $\int_0^\tau\omega_p(t)dt=\pi$. Through the Hamiltonian Eq.~(\ref{H1}), we can write the evolution operator of the Sagnac interferometer
\begin{eqnarray}
\label{U}
U(\tau)&=&\bigotimes\limits_{k=1}^{N}U_k(\tau)\nonumber\\
&=&\bigotimes\limits_{k=1}^{N}e^{-i\omega a_k^\dagger a_k\tau}
e^{i\Phi_k(\omega,\Omega,\tau)}D_k[\eta_k(\omega,\Omega,\tau)],
\end{eqnarray}
where
\begin{align}
\label{Phik}\begin{split}
\Phi_k(\omega,\Omega,\tau)=&\int_{0}^{\tau}\int_{0}^{t_1}f_k(\omega,\Omega,t_1)
f_k(\omega,\Omega,t_2)\\
&\cdot\sin(\omega(t_1-t_2))dt_2dt_1,\\
\eta_k(\omega,\Omega,\tau)=&-\int_{0}^{\tau}f_k(\omega,\Omega,t)e^{i\omega t}dt,\\
f_k(\omega,\Omega,t)=&\sqrt{\dfrac{m\omega}{2\hbar}}R\left(\Omega+\sigma_z^{(k)}\omega_p(t)\right).
\end{split}
\end{align}
and $D_k[\eta]=e^{\eta a_k^\dagger-\eta^\ast a_k}$ refers to the displacement operator. \\
\par
\noindent For a quantum estimation with multiple parameters $(\theta_1,\theta_2,\cdots)$ in a pure state $|\psi\rangle$, its precision is scaled by the Cram$\rm \acute{e}$r-Rao bound
\begin{eqnarray}
\label{CR1}
\mathcal{C}\geqslant\frac{1}{\nu}\mathcal{F}^{-1}.
\end{eqnarray}
Where $\nu$ is the times of experiments, $\mathcal{C}$ is the estimation-error covariance matrix of the parameters with elements $\mathcal{C}_{ij}=\rm{Cov}(\theta_i,\theta_j)=\langle\psi|\frac{1}{2}(\theta_i\theta_j+\theta_j\theta_i)|\psi\rangle-
\langle\psi|\theta_i|\psi\rangle\langle\psi|\theta_j|\psi\rangle$, $\mathcal{F}$ is the quantum Fisher information matrix(QFIM) that is defined as $\mathcal{F}_{ij}=4\rm{Re}(\langle\partial_{\theta_i}\psi|\partial_{\theta_j}\psi\rangle-
\langle\partial_{\theta_i}\psi|\psi\rangle\langle\psi|\partial_{\theta_j}\psi\rangle)$ and $\mathcal{F}^{-1}$ refers to the inverse matrix of $\mathcal{F}$ (here we only consider the full rank matrix $\mathcal{F}$ case).
If the final state is generated by imposing an unitary operation to the initial state, i.e., $|\psi\rangle=U|\psi_0\rangle$, the QFIM can be simplified as $\mathcal{F}_{ij}=4\rm{Cov}(\mathcal{H}_{\theta_i},\mathcal{H}_{\theta_j})=4(\langle\psi_0|
\frac{1}{2}(\mathcal{H}_{\theta_i}\mathcal{H}_{\theta_j}+\mathcal{H}_{\theta_j}\mathcal{H}_{\theta_i})|\psi_0\rangle-\langle\psi_0|
\mathcal{H}_{\theta_i}|\psi_0\rangle\langle\psi_0|\mathcal{H}_{\theta_j}|\psi_0\rangle)$  with $\mathcal{H}_{\theta_i}=i(\partial_{\theta_i}U^\dagger)U$ is the generator of $U$\cite{Liu2013}. Then in our scenario of two-parameter($\omega$,$\Omega$) measurement, the Cram$\rm \acute{e}$r-Rao bound is
\begin{equation}
\label{CR2}
\left(\!\!\!\begin{array}{cc}
\delta^2\omega & \!\!\!\rm{Cov}(\omega,\Omega)\\
\rm{Cov}(\Omega,\omega) &\!\!\! \delta^2\Omega\\
\end{array}\!\!\!\right)\geqslant
\frac{1}{4}\left(\!\!\!\begin{array}{cc}
\Delta^2\mathcal{H}_\omega & \!\!\! \rm{Cov}(\mathcal{H}_\omega,\mathcal{H}_\Omega)\\
\rm{Cov}(\mathcal{H}_\Omega,\mathcal{H}_\omega) & \!\!\! \Delta^2\mathcal{H}_\Omega\\
\end{array}\!\!\!\right)^{-1},
\end{equation}\\
here we have set $\nu=1$ for simplicity and $\delta^2\omega(\Omega)$ is the variance of parameter $\omega(\Omega)$ while  $\Delta^2\mathcal{H}_{\omega(\Omega)}=\langle\psi_0|\mathcal{H}_{\omega(\Omega)}^2|\psi_0\rangle-
\langle\psi_0|\mathcal{H}_{\omega(\Omega)}|\psi_0\rangle^2$ denotes the measurement variance of $\mathcal{H}_{\omega(\Omega)}$ in the input state. Note that the symbol $\geqslant$ in Eq.~(\ref{CR1}) means for any vector $\textbf{x}=(x_1,x_2,\cdots,)^\mathsf{T}$, Eq.~(\ref{CR1}) should satisfy $\textbf{x}^\mathsf{T}\mathcal{C}\textbf{x}\geqslant \textbf{x}^\mathsf{T}\mathcal{F}^{-1}\textbf{x}$, let $\textbf{x}=(0,\cdots,0,1,0,\cdots)^\mathsf{T}$, and we will obtain $\mathcal{C}_{ii}\geqslant\mathcal{F}^{-1}_{ii}$. Using this in Eq.~(\ref{CR2}), we get
\begin{align}
\label{deltaomega}\begin{split}
\delta^2\omega\geqslant&\dfrac{1}{4}\dfrac{\Delta^2\mathcal{H}_{\Omega}}{\Delta^2
\mathcal{H}_{\omega}\Delta^2\mathcal{H}_{\Omega}-\left|{\rm Cov}(\mathcal{H}_{\omega},\mathcal{H}_{\Omega})\right|^2},\\
\delta^2\Omega\geqslant&\dfrac{1}{4}\dfrac{\Delta^2\mathcal{H}_{\omega}}{\Delta^2
\mathcal{H}_{\omega}\Delta^2\mathcal{H}_{\Omega}-\left|{\rm Cov}(\mathcal{H}_{\omega},\mathcal{H}_{\Omega})\right|^2},\end{split}
\end{align}
where the two equality signs hold simultaneously if and only if\cite{Liu2015}
\begin{eqnarray}
\label{condition1}
\langle\psi_0|\left[\mathcal{H}_{\omega},\mathcal{H}_{\Omega}\right]|\psi_0\rangle=0.
\end{eqnarray}
The variances in the last equations reflect the absolute value of the fluctuate of the measured parameters, which is not a very suitable criterion to scale the measurement precision as the same fluctuation of different sample values usually signify quite different stability. We think the relative variances, which is defined as $\delta^2\omega(\Omega)_{\rm r}=\delta^2\omega(\Omega)/{\omega^2_0(\Omega^2_0)}$, with $\omega_0(\Omega_0)$ is the selected true value of the parameter, is more appropriate, so in the following parts, we would calculate and discuss the relative variances rather than the absolute variances.\\

\vspace*{-1.5ex}
\section{Calculating The Measurement Precision}
\vspace*{-1.5ex}
\label{calculating}
Through Eq.~(\ref{U}), $\mathcal{H}_{\omega(\Omega)}$ can be expressed as
\begin{eqnarray}
\label{hohO}
\mathcal{H}_{\omega}=\sum\limits_{k=1}^{N}\mathcal{H}_\omega^{(k)},~~~~~
\mathcal{H}_{\Omega}=\sum\limits_{k=1}^{N}\mathcal{H}_\Omega^{(k)},
\end{eqnarray}
with $\mathcal{H}_{\omega(\Omega)}^{(k)}=i(\partial_{\omega(\Omega)}U_k(\tau)^\dagger)U_k(\tau)$ is the single particle generator operator.
Then the elements of the QFIM is given by
\begin{align}
\label{FIM1}
\Delta^2\mathcal{H}_{\omega(\Omega)}=&\sum\limits_{k=1}^N\Delta^2\mathcal{H}_{\omega
(\Omega)}^{(k)}+\sum\limits_{k_1\ne k_2}^{N}{\rm Cov}(\mathcal{H}_{\omega(\Omega)}^{(k_1)},
\mathcal{H}_{\omega(\Omega)}^{(k_2)}),\nonumber\\
=&N\Delta^2\mathcal{H}_{\omega(\Omega)}^{(k)}\notag\\
&+(N^2-N){\rm Cov}(\mathcal{H}_{\omega(\Omega)}^{(k_1)},\mathcal{H}_{\omega(\Omega)}^{(k_2)}),
\end{align}
\begin{align}
\label{FIM2}
{\rm Cov}(\!\mathcal{H}_{\omega},\!\mathcal{H}_{\Omega}\!)\!=&{\rm Cov}(\mathcal{H}_{\Omega},\mathcal{H}_{\omega})\nonumber\\
=&\sum\limits_{k=1}^{N}{\rm Cov}
(\mathcal{H}_{\omega}^{(k)},\mathcal{H}_{\Omega}^{(k)})\!+\!\!\!\sum\limits_{k_1\ne k_2}^{N}\!\!{\rm Cov}
(\mathcal{H}_{\omega}^{(k_1)},\mathcal{H}_{\Omega}^{(k_2)})\nonumber\\
=&N{\rm Cov}(\mathcal{H}_{\omega}^{(k)},\mathcal{H}_{\Omega}^{(k)})\notag\\
&+(N^2-N){\rm Cov}(\mathcal{H}_{\omega}^{(k_1)},\mathcal{H}_{\Omega}^{(k_2)}),
\end{align}
where the second step in Eq.~(\ref{FIM1}) and the third step in Eq.~(\ref{FIM2}) have considered the fact that as each $\mathcal{H}^{(k)}_{\omega(\Omega)}$  has the same position in our system, all variances or covariances of the generator operators should be equal in the input state. Thereby the constraint condition in Eq.~(\ref{condition1}) becomes
\begin{align}
\label{condition2}
\langle\psi_0|\sum\limits_{k=0}^N[\mathcal{H}_\omega^{(k)},\mathcal{H}_\Omega^{(k)}]
|\psi_0\rangle=0~\Leftrightarrow~\langle\psi_0|[\mathcal{H}_\omega^{(k)},\mathcal{H}_\Omega^{(k)}]
|\psi_0\rangle=0.
\end{align}
According to the expression of the operator $U_k(\tau)$ in Eq.~(\ref{U}), $\mathcal{H}_{\omega(\Omega)}^k$ can be further expressed as
\begin{eqnarray}
\label{ho1}
\begin{split}
\mathcal{H}_\omega^{(k)}=&\left[(i\partial_\omega\eta_k^\ast-\tau\eta_k^\ast)a_k+h.c.)\right]
+i\dfrac{1}{2}(\eta_k\partial_\omega\eta_k^\ast-h.c.)\\
&+\partial_\omega\Phi_k-\tau a_k^\dagger a-\tau|\eta_k|^2,\\
\mathcal{H}_\Omega^{(k)}=&i(\partial_\Omega\eta_k^\ast a_k-h.c.)+i\dfrac{1}{2}(\eta_k\partial_\Omega\eta_k^\ast-h.c.)+\partial_\Omega\Phi_k.
\end{split}
\end{eqnarray}
Here we have let $\eta_k=\eta_k(\omega,\Omega,\tau)$, $\Phi_k=\Phi_k(\omega,\Omega,\tau)$ in Eqs.~(\ref{ho1}) for cleanness. Inserting Eqs.~(\ref{Phik}) into the last equalities and ignoring the constant terms, yields
\begin{align}
\mathcal{H}_\omega^{(k)}=&(K_1 a_k+K_1^\ast a_k^\dagger)-(K_2 a_k+K_2^\ast a_k^\dagger)\sigma_z^{(k)}\nonumber\\
&+\lambda\sigma_z^{(k)}-\tau a_k^\dagger a_k,\label{ho2}\\
\mathcal{H}_\Omega^{(k)}=&(\delta_1 a_k+\delta_1^\ast a_k^\dagger)+\delta_2\sigma_z^{(k)},\label{hO2}
\end{align}
where
\begin{align}
\label{K1}\begin{split}
K_1=&\mu\sqrt{\omega}\Omega\left[\left(\tau-i\dfrac{1}{2\omega}\right)q^\ast(\omega,\tau)
-i\partial_\omega q^\ast(\omega,\tau)\right],\\
K_2=&\mu\sqrt{\omega}\left[\left(i\dfrac{1}{2\omega}-\tau\right)p^\ast(\omega,\tau)
+i\partial_\omega p^\ast(\omega,\tau)\right],\\
\lambda=&\mu^2\Omega\left\{\dfrac{1}{\omega}\int_{0}^{\tau}\omega_p(t)[\cos\omega(t-\tau)
-\cos(\omega t)]dt\right.\\
&\left.+2\int_0^{\tau}\omega_p(t)(t-\tau)\sin(\omega t)dt\right\},\\
\delta_1=&-i\dfrac{2\mu}{\sqrt{\omega}}\sin\left(\dfrac{\omega\tau}{2}\right)e^{-i\frac{\omega}{2}\tau},\\
\delta_2=&2\pi\mu^2\left(1-\dfrac{1}{\pi}\int_0^\tau\omega_p(t)\cos[\omega(\tau-t)]dt\right),
\end{split}
\end{align}
with $\mu=\sqrt{\dfrac{m}{2\hbar}}R$, $p(\omega,\tau)=\int_0^\tau\omega_p(t)e^{i\omega t}dt$ and $q(\omega,t)=\int_0^\tau e^{i\omega t}dt$.
The commutation relation between $\mathcal{H}_\omega^{(k)}$ and $\mathcal{H}_\Omega^{(k)}$ is $[\mathcal{H}_\omega^{(k)},\mathcal{H}_\Omega^{(k)}]=(K_1\delta_1^\ast-K_1^\ast\delta_1)
+(\delta_1K_2^\ast-\delta_1^\ast K_2)\sigma_z^{(k)}+(\delta_1a_k-\delta_1^\ast a_k^\dagger)\tau$. Substituting Eqs.~(\ref{K1}) and considering the fact that spin up and down particles in the input state is half to half that leads to $\langle\psi_0|\sigma_z^{(k)}|\psi_0\rangle=0$, the constraint condition of Eq.~(\ref{condition2}) can be simplified as
\begin{eqnarray}
\label{Condition1}
\sin\left(\frac{\omega_0\tau}{2}\right)=0~~~\Rightarrow~~~\tau=\kappa\cdot\frac{2\pi}{\omega_0},
~~~\kappa=1,2,\cdots,
\end{eqnarray}
or
\begin{eqnarray}
\label{Condition2}
\frac{\mu\Omega_0}{\sqrt{\omega_0}}\sin\left(\frac{\omega_0\tau}{2}\right)={\rm Re}\left(e^{-i\frac{\omega_0\tau}{2}}\langle\psi_0|a_k|\psi_0\rangle\right)\neq&0,
\end{eqnarray}
here ${\rm Re}(\cdot)$ means the real part of a complex number. We also have replaced $\omega(\Omega)$ with the selected true value of $\omega_0(\Omega_0)$ in the last equations. Up to now we have prepared everything well to calculate the measurement precision. The calculation process can be reduced as follow: Combining Eqs.~(\ref{ho2})$\sim$(\ref{hO2}),~(\ref{FIM1})$\sim$(\ref{FIM2}), we figure out the general expression of the QFIM, again combining Eqs.~(\ref{deltaomega}), we would get the lower limits of the measured parameters' relative variances, while Eqs.~(\ref{Condition1})$\sim$(\ref{Condition2}) enable the two lower bounds can be achieved simultaneously. Next we would analyse the measurement results by considering the input state.\\
\par

\vspace*{-1.5ex}
\section{General Analysis Of The Measurement Precision}
\vspace*{-1.5ex}
\label{general analysis}
We first calculate the terms in Eqs.~(\ref{FIM1})$\sim$(\ref{FIM2}) from Eqs.~(\ref{ho2})$\sim$(\ref{hO2}) under a general input state in Eq.~(\ref{input1}). We find
\begin{eqnarray}
\label{general1}\begin{split}
\Delta\mathcal{H}^2_\omega=AN+BN^2,~~~~~~~~\Delta\mathcal{H}^2_\Omega=CN+DN^2,\\
\Delta\mathcal{H}^2_\omega\Delta\mathcal{H}^2_\Omega-
|{\rm Cov}(\mathcal{H}_\omega,\mathcal{H}_\Omega)|^2=EN^2+FN^3.\end{split}
\end{eqnarray}
Where $A,B,C,D,E,F$ are real prefactors that are related to the selected true value $\omega_0(\Omega_0)$, the evolution time $\tau$ and the input state. The concrete expressions of theses prefactors, especially $A,E,F$, are extremely long, we don't plan to list them in this paper, we just analyse them. From Eqs.~(\ref{deltaomega}), it's clearly that if we want the measurement precision limit of $\omega(\Omega)$ to reach the HL when $N\gg1$, we should let $B(D)=0$. The condition that makes $B(D)=0$ is
\begin{align}
\label{B=0}
B=0~\Leftrightarrow~&2{\rm Re}\left(K_1\langle\psi_0|a_k\sigma_z^{(k)}|\psi_0\rangle-
K_2\langle\psi_0|a_k|\psi_0\rangle\right)+\lambda\notag\\
&=\tau\langle\psi_0|a_k^\dagger a_k\sigma_z^{(k)}|\psi_0\rangle,\\
D=0~\Leftrightarrow&~\frac{\delta_2}{2}+{\rm Re}\left(\delta_1\langle\psi_0|a_k\sigma_z^{(k)}|\psi_0\rangle\right)=0.\label{D=0}
\end{align}
Here $K_1,K_2,\lambda,\sigma_1,\sigma_2$ are seen in Eqs.~(\ref{K1}). The right side of Eq.~(\ref{B=0}) can be denoted as the mean energy difference between spin up and down particles as $\langle\psi_0|a_k^\dagger a_k\sigma_z^{(k)}|\psi_0|\rangle=1/2(\langle\psi_\uparrow|a_k^\dagger a_k|\psi_\uparrow\rangle-\langle\psi_\downarrow|a_k^\dagger a_k|\psi_\downarrow\rangle)=1/(2\hbar\omega_0)(\bar{E}_k^{\uparrow}-\bar{E}_k^{\downarrow})
=1/(2N\hbar\omega_0)(\bar{E}^{\uparrow}-\bar{E}^{\downarrow})$. Unfortunately, calculating shows if conditions in Eqs.~(\ref{B=0}) and ~(\ref{D=0}) are both met, that will lead $F$ be equal to zero as well. Which means the optimal measuring results of the two parameters cannot achieve the HL simultaneously in current scenario. But if only one of the last two equations is satisfied, for example, $B$ or $D$ is equal to zero, there would be $A\cdot D=F$ or $B\cdot C=F$, hence for large particle number cases, namely, $N\gg1$, we would have
\begin{eqnarray}\begin{split}
\label{B=02}
&B=0\Rightarrow AD=F\\
&\Rightarrow\delta^2\omega_{\rm r}\sim\frac{1}{N}\cdot
\frac{1}{4\omega_0^2A},~\delta^2\Omega_{\rm r}\sim\frac{1}{N^2}\cdot\frac{1}{4\Omega_0^2D}.\end{split}
\end{eqnarray}
\begin{eqnarray}\begin{split}
\label{D=02}
&D=0\Rightarrow BC=F\\
&\Rightarrow\delta^2\omega_{\rm r}\sim\frac{1}{N^2}\cdot
\frac{1}{4\omega_0^2B},~\delta^2\Omega_{\rm r}\sim\frac{1}{N}\cdot\frac{1}{4\Omega_0^2C}.\end{split}
\end{eqnarray}
In other words, with the input state in Eq.~(\ref{input1}) and the Sagnac System depicted in Fig.1, at most one measurement result can reach the HL while the other would be scaled by the SQL.
Such as it is, we may be able to choose which one be the HL. We have mentioned before that if we want the two relative variances reach their lower limits at the same time, a constraint condition in Eq.~(\ref{Condition1}) or (\ref{Condition2}) should be satisfied. Now we consider the constraint condition.\\
We call Eq.~(\ref{Condition1}) Condition ${\rm \RNum{1}}$, Eq.~(\ref{Condition2}) Condition ${\rm \RNum{2}}$. We would do some further analysis about the measurement precision under these two conditions. For the sake of simplicity, we set the relative rotating angular velocity $\omega_p(t)$ a constant function. This is a feasible choice in actual experiments as $\omega_p(t)$ is constant means the potential well rotates uniformly around the Sagnac apparatus, which is easier to be operated and has higher stability than the variable rotation system.

\vspace*{-1.5ex}
\section{Constraint Condition \RNum{1}}
\vspace*{-1.5ex}
\label{CC1}
Condition Eq.~(\ref{Condition1}) determines the evolution time $\tau$ with a series of discrete values that leads $\delta_1$ to $0$. Then $\omega_p=\omega_0/(2\kappa)$, with $\kappa$ any positive integer. Parameters in Eqs.(\ref{K1}) are also determined
\begin{eqnarray}
\begin{split}
\label{K1delta2}
K_1&=-i\frac{2\mu\pi\kappa\Omega_0}{\omega_0^{3/2}},~~~~~K_2=i\frac{\mu\pi}{\sqrt{\omega_0}}\\
\lambda&=-\frac{2\mu^2\pi\Omega_0}{\omega_0},~~~~~~~~\delta_2=2\mu^2\pi.
\end{split}
\end{eqnarray}
Here $\mu=\sqrt{m/(2\hbar)}R$, represents the feature of the Sagnac interferometer, $\omega_0(\Omega_0)$ is the selected true value of $\omega(\Omega)$. Such a simplification makes $C,E$ in Eqs.~(\ref{general1}) equal to zero, meanwhile $D=\delta_2^2, A/F=1/\delta_2^2$. As $D$ is not zero, the minimum uncertainty of $\omega$ cannot reach the HL. Therefore, we should let $B=0$ in this part. From Eq.~(\ref{B=02}), we find the lower limit of the relative variance $\delta\omega(\Omega)_{\rm r}$ is
\begin{eqnarray}
\label{con1omega}
\delta^2\omega_{{\rm r}}=\frac{1}{N}\cdot\frac{1}{4\omega_0^2A},
\end{eqnarray}
\begin{eqnarray}
\label{con1Omega}
\delta^2\Omega_{{\rm r}}=\frac{1}{N^2}\cdot\frac{1}{4\Omega_0^2\delta_2^2}
=\frac{1}{N^2}\cdot\frac{1}{16\pi^2\mu^4\Omega_0^2},
\end{eqnarray}
Obviously the measurement precision of $\omega$ is bounded by the SQL while the one of $\Omega$ can reach the HL. Besides, this HL item is the same as the one with single parameter estimation\cite{Luo2017}. Generally speaking, compared with single parameter estimation, multi-parameter estimation has more constraint conditions and hence the measurement precision is usually lower, but here we get the same result, which means our measurement scenario is advisable to maintain measurement precision that is obtained by single parameter measurement scenario. Additionally, the result in Eq.~(\ref{con1Omega}) is determined only by the Sagnac interferometer and the particle mass while has no concern with the trapping well, that is to say, whatever trapping potential and input state are chosen, we would always get the same HL result when measuring $\Omega$. Such a stability may quite useful in practical experiments. The input state here only affects the measurement precision of $\omega$, whose influence is reflected in the value of the prefactor $A$ that larger value of $A$ leads to more precise result. We would post some special states as examples to calculate the lower limit of $\delta^2\omega_{\rm r}$ and compare which is better.

\begin{figure*}
\centerline{\includegraphics[width=7.5in,clip]{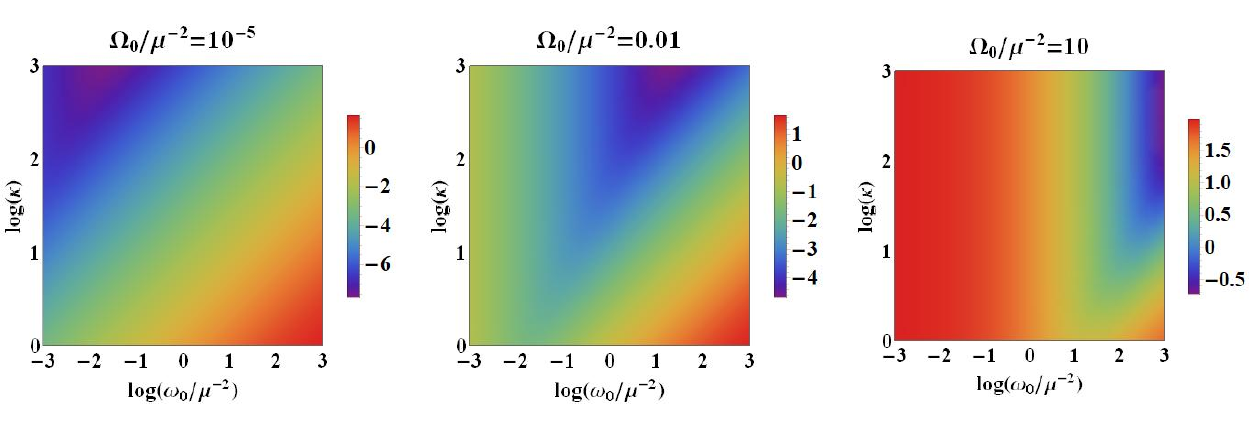}}
\caption{The distribution of the ${\rm log}$ scaling ratio between the minimum relevant variances of $\omega$ in the coherent state and Fock state cases. We set $\mu^{-2}$ as the unit of the frequency and let the total energy levels,i.e., $n_1+n_2$ or $|\alpha_1|^2+|\alpha_2|^2$ is 100. $\kappa$ represents the evolution time. Points with value larger than zero means Fock state is better, otherwise the coherent state case gets more precise measurement result.}
\label{fig:GeoPhase}
\end{figure*}

\subsection{Fock State}
Assume the particles' position states are eigenstates of the harmonic trapping potential. Without loss of generality, we set $|\psi_\uparrow\rangle=|n_1\rangle$, $|\psi_\downarrow\rangle=|n_2\rangle$. Then the input state is $|\psi_0\rangle=1/\sqrt{2}\left(\bigotimes\limits_{k=1}^{N}|n_1,\uparrow\rangle_k
+\bigotimes\limits_{k=1}^{N}|n_2,\downarrow\rangle_k\right)$. The condition in Eq.~(\ref{B=0}) therefore can be transformed as $\hat{E}^{\downarrow}-\hat{E}^{\uparrow}=N\hbar\omega_0(n_2-n_1)/2=N\omega_0\Omega_0\hbar\mu^2/\kappa$, where $\hat{E}^{\uparrow(\downarrow)}$ refers to the total average trapping energy of the spin up(down) particles. That means if we want the best measurement precision of $\Omega$ to reach the HL, the spin down particles in the trapping well should have higher energy than the spin up ones with a certain energy gap, which as well determines the energy level difference between the two types of particles that $n_2-n_1=2\Omega_0\mu^2/\kappa$. The minimum relative variances of $\omega$ now yields
\begin{align}
\label{forkomega}
(\delta^2\omega_{\rm r})_{\rm m}=\frac{1}{N}\frac{1}{4\mu^2\pi^2[(n+1)
(\omega_0+4\kappa^2\Omega_0^2/\omega_0)+8\mu^2\Omega_0^2]}.
\end{align}
Where $n=n_1+n_2$ is the gross energy level that satisfies $n\geqslant 2\Omega_0\mu^2/\kappa$ with equality only holds for $n_1=0$, that is, the spin up particles must be prepared in the ground state. We can see the true value of $\omega(\Omega)$, the total energy levels of the particles and the evolution time would affect the measurement precision. For a given Sagnac metrology system, that means $\mu$ is fixed, to improve the measurement precision, the value of $\Omega_0$, the evolution time and the gross trapping energy level of the particles should be as large as possible, while the value of $\omega_0$ should be selected as far away as possible from $2\kappa\Omega_0$.

\vspace*{-1.5ex}
\subsection{Coherent State}
\vspace*{-1.5ex}
Here we calculate the measurement result of $\omega$ when the motional states are coherence states, i.e.,$|\psi_\uparrow\rangle=|\alpha_1\rangle$, $|\psi_\downarrow\rangle=|\alpha_2\rangle$. Let $\alpha_1=r_1e^{i\theta_1}, \alpha_2=r_2e^{i\theta_2}$, with $r_1, r_2$ positive real numbers, $\theta_1,\theta_2\in[-\pi,\pi]$. We find the minimum relative variance of $\omega$ is
\begin{eqnarray}
\label{alphaomega}
(\delta^2\omega_{\rm r})_{\rm m}=\frac{1}{N}\frac{1}{4\pi^2\left[2\kappa r_1+\mu(\sqrt{\omega_0}+2\kappa\Omega/\sqrt{\omega_0})\right]^2}.
\end{eqnarray}
To obtain this equation, the input coherent states should satisfy
\begin{align}
\label{alphacondition}
\theta_1\!=\!-\dfrac{\pi}{2},\begin{cases}\theta_2\!=\!\dfrac{\pi}{2},~ r_2\!=\!r_1\!+\!\dfrac{2\mu\Omega_0}{\sqrt{\omega_0}},\!\!\!&\rm{if} ~~\omega_0 \!> \!2\kappa\Omega_0\\
\theta_2\!=\!-\dfrac{\pi}{2},~r_2\!=\!r_1\!+\!\dfrac{\mu\sqrt{\omega_0}}{\kappa}.\!\!\!&\rm{if} ~~\omega_0\!<\!2\kappa\Omega_0\end{cases}
\end{align}
Let $n=r_1^2+r_2^2$, indicates the total energy level of one pair anti-spin particles, then Eq.~(\ref{alphaomega}) can be rewritten as
\begin{align}
\label{alphaomega2}
(\delta^2\omega_{\rm r})_{\rm m}\!=&\!\dfrac{1}{N}\cdot\dfrac{1}{16\pi^2\kappa^2}\cdot\nonumber\\
&\!\!\!\!\!\begin{cases}
\dfrac{1}{\left(\!\sqrt{(n\!-\!2\mu^2\Omega_0^2/\omega_0)/2}\!+\!\dfrac{\mu\sqrt{\omega_0}}{2\kappa}\!\right)^2},\!\!\!&\rm{if}~
\omega_0\!>\!2\kappa\Omega_0\\
\dfrac{1}{\left(\!\sqrt{(n\!-\!\mu^2\omega_0/(2\kappa^2))/2}\!+\!\dfrac{\mu\Omega_0}{\sqrt{\omega_0}}\!\right)^2}.\!\!\!&\rm{if}~
\omega_0\!<\!2\kappa\Omega_0\end{cases}
\end{align}
Just like the previous Fock state example, here $n$ as well has a lower bound, i.e., $n\geqslant\min[2\mu^2\Omega_0^2/\omega_0,\mu^2\omega_0/(2\kappa^2)]$. From Eq.~(\ref{alphaomega2}) we can see larger $n$ and $\kappa$ would lead to better measurement precision,nwhich is the same as the Fock state example. But the influence of the selected true value $\omega_0(\Omega_0)$ here is not so obviously.\\
We would like to compare which of the two examples is more suitable for measuring $\omega$, as seen in Fig.2. We find the effect of the input resource is strongly influenced by $\Omega_0$. When $\Omega_0$ is large, see the right picture, almost all (indeed all if $\Omega_0$ is larger) the points in the picture has a value larger than zero, which means the measurement result obtained by the Fock state is almost always better than that achieved by the coherent state. The right picture also shows such superiority of Fock state is less influenced by the evolution time and strengthened when the selected $\omega_0$ is smaller. On the contrary, when $\Omega_0$ is small, see the left and middle pictures, the coherent state will be a better choice when the evolution time is large or the true value of $\omega_0$ is small. However, no matter how small the $\Omega_0$ is, the values of the points in the bottom-right area (with large $\omega_0$ and low $\kappa$) of the pictures are still larger than zero, means the Fock state still gets more precise result if we measurement $\omega$ around a large true value with short evolution time.\\
\par

\vspace*{-1.5ex}
\section{Constraint Condition \RNum{2}}
\vspace*{-1.5ex}
\label{CC2}
Assume $\langle\psi_0|a_k|\psi_0\rangle=x+iy$ with $x,y$ are real numbers. Then constraint condition of Eq.~(\ref{Condition2}) yields
\begin{eqnarray}
\label{Con2}
\frac{\mu\Omega_0}{\sqrt{\omega_0}}\sin\left(\frac{\omega_0\tau}{2}\right)\!=\!x\cos\left(
\frac{\omega_0\tau}{2}\right)\!+\!y\sin\left(\frac{\omega_0\tau}{2}\right)\!\ne\!0.
\end{eqnarray}
It's hard to calculate the measurement precision if the evolution time $\tau$ is correlated with the input state. In this part, we only discuss a relatively simple situation that $\cos(\omega_0\tau/2)=0$, then the constraint condition of Eq.~(\ref{Con2}) becomes
\begin{eqnarray}
\tau=\frac{\pi}{\omega_0}(2\kappa+1),~~~~~~~~~~~y=\frac{\mu\Omega_0}{\sqrt{\omega_0}}.
\end{eqnarray}
Where $\kappa$ is any arbitrary natural number. Just like the previous section, here we let the evolution time is independent with the input state and is determined only by $\omega_0$, so $\omega_p=\omega_0/(2\kappa+1)$ and the parameters in Eqs.~(\ref{K1}) are
\begin{eqnarray}\begin{split}
K_1=&\dfrac{\Omega_0\mu}{\omega_0^{3/2}}\left[1-i(2\kappa+1)\pi\right],\\
K_2=&\dfrac{\mu}{\sqrt{\omega_0}}\left(-\dfrac{1}{2\kappa+1}+i\pi\right),\\
\lambda=&-\dfrac{2\mu^2\pi\Omega_0}{\omega_0},~~~~~\delta_1=-\dfrac{2\mu}{\sqrt{\omega_0}},~~~~~
\delta_2=2\mu^2\pi.\end{split}
\end{eqnarray}
Unlike the previous part, here the prefactor $B$ or $D$ is not necessarily equal to zero. If $B=0\&D\ne0$, the measurement results would have the form as Eq.~(\ref{B=02}) while if $D=0\&B\ne0$, the results will be the form of Eq.~(\ref{D=02}). So each of the two parameter's minimum uncertainty may reach the HL. The concrete general expressions of the prefactors $A\sim D$ is still too complex, for the sake of simplicity, we shall explore the measurement precision  just through some examples. We choose the coherent state, i.e.,$|\psi_\uparrow\rangle=|\alpha_1\rangle$, $|\psi_\downarrow\rangle=|\alpha_2\rangle$ as the position input state in the following examples. Then the prefactors of $A\sim D$ is simplified as
\begin{eqnarray}
\label{A-D}
\begin{split}
A=&\dfrac{1}{2}\left(|\alpha_1\tau+K_2^\ast-K_1^\ast|^2+|\alpha_2\tau-K_1^\ast-K_2^\ast|^2\right),\\
B=&\Big[\lambda+{\rm Re}(K_1(\alpha_1-\alpha_2))-{\rm Re}(K_2(\alpha_1+\alpha_2))\Big.\\
&\Big.-\dfrac{1}{2}\tau(|\alpha_1|^2-|\alpha_2|^2)\Big]^2,\\
C=&|\delta_1|^2,\\
D=&\left(\delta_2+{\rm Re}(\delta_1(\alpha_1-\alpha_2))\right)^2.\end{split}
\end{eqnarray}
We also set $\alpha_1=x_1+iy_1, \alpha_2=x_2+iy_2$. Consider the fact that $\langle\psi_0|a_k|\psi_0\rangle=x+iy=1/2[x_1+x_2+i(y_1+y_2)]$, the constraint conditions about the input state becomes
\begin{eqnarray}
y_1+y_2=\frac{2\mu\Omega_0}{\sqrt{\omega_0}}.
\end{eqnarray}
We would further analyse the ultimate measurement precisions with the condition $B=0$ or $D=0$.


\begin{figure}
\centering
\includegraphics[width=3.7 in]{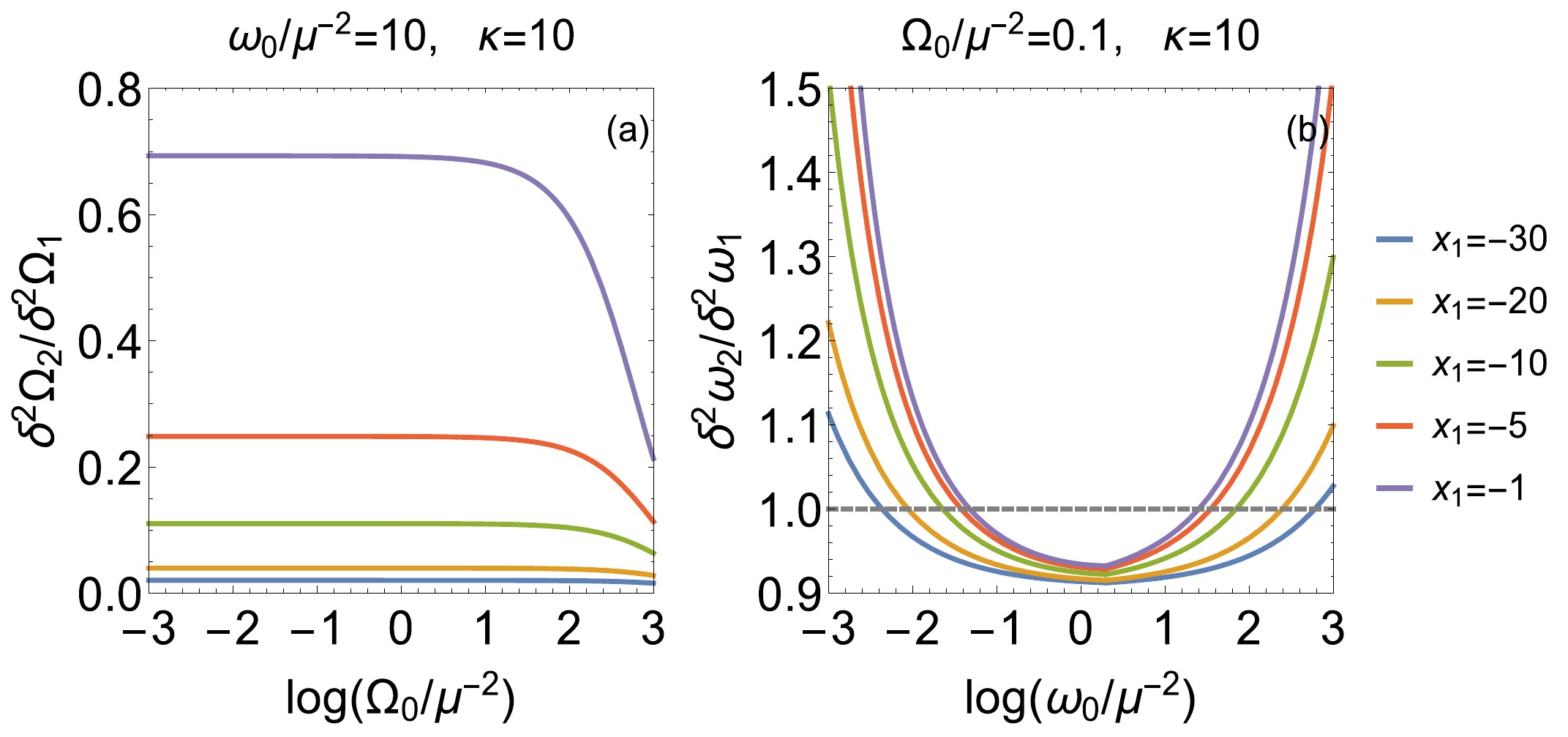}
\protect\caption{The ratio of $\delta^2\Omega(\omega)$ with coherent input state in Condition ${\rm\RNum{2}}$ to that in Condition ${\rm\RNum{1}}$
versus the true value of $\Omega_0(\omega_0)$ with different negative $x_1$. We set coherent states in both conditions have the same trapping energy, i.e., $x_1^2+x_2^2+y_1^2+y_2^2=r_1^2+r_2^2$ and let $y_1=10$. In (a), all lines lie in the below-one area and the line with smaller $x_1$ is positioned lower. In (b), each line is similar to an upward opening parabola with the values of the bottom part are smaller than one, such part would expand when $x_1$ becomes smaller.      }
\label{Fig:3}
\end{figure}


\subsection{$B=0~~~\textbf{and}~~~D\ne0$}
First consider the case of $B=0\&D\ne0$. $B=0$ leads to a constraint relationship between $x_1$ and $x_2$
\begin{align}
x_2=&\dfrac{1}{\pi(2\kappa+1)^2\sqrt\omega_0}\Big\{\mu(-\omega_0+\Omega_0+2\kappa\Omega_0)\Big.\notag\\
&\Big.+\left[\pi^2x_1^2(2\kappa+1)^4\omega_0+\mu^2(-\omega_0+\Omega_0+2\kappa\Omega_0)^2\right.\notag\\
&\left.-2\pi x_1(2\kappa+1)^2\mu\sqrt\omega_0(\omega_0+\Omega_0+2\kappa\Omega_0)\right]^{\frac{1}{2}}\Big\}.
\end{align}
Therefore, with each of the two is settled, $D$ would be identified.  Note that $\delta_2$ here is the same as the one in Eq.~(\ref{con1Omega}) and $\delta_1$ is a negative number, so the prefactor $D$ may be larger than the one in Condition ${\rm\RNum{1}}$ section and hence we would get a more precise measurement result of $\Omega$ if $x_1-x_2$ is also negative, which can be testified always true when $x_1$ is less than zero. Fig.3(a) intuitively shows some cases that the coherent input state with $x_1<0$ indeed performs better when measuring $\Omega$ here than the previous one, no matter what the true value of $\Omega_0$ be selected. Besides, the comparative measurement advantage becomes more obvious when the absolute value of $x_1$ is larger. \\
We also do a comparison of the measurement results of $\omega$ here and the one in Eq.~(\ref{alphaomega2}), as seen in Fig.3(b). Under the condition that $\Omega_0=0.1\mu^{-2}$, $\kappa=10$, the measurement precision is better in this part only if the true value is selected not very far away from $\mu^{-2}$, and the range of such proper $\omega_0$ would expand when the absolute value of $x_1$ becomes larger. On the contrary, if $\omega_0$ is small enough or large enough, coherent state in Condition ${\rm\RNum{1}}$ would have a more precise measurement result.\\
Combining these two pictures, we can say in some special cases, such as  $\omega_0\approx\mu^{-2}$, $\Omega_0\approx 0.1\mu^{-2}$ and short evolution time, $\tau\sim21\pi/\omega_0$, the coherent state input resource in Condition ${\rm\RNum{2}}$ with negative $x_1$ would get more precise measurement results of both parameters than the same trapping energy coherent state in Condition ${\rm\RNum{1}}$. As the measurement result of $\Omega$ in Condition ${\rm\RNum{1}}$ is the same as that in single parameter estimation scenario, we would say with a Sagnac interferometer, at least in some special cases, measuring parameters in multiparameter schemes, the precision can surpass the one in individual parameter measurement schemes. \\
\par

\subsection{$D=0~~~\textbf{and}~~~B\ne0$}
When $B\ne0\&D=0$, that will lead to $x_2=x_1-\mu\pi\sqrt{\omega_0}$ and the prefactors of $B,C$ here can be written as
\begin{align}
B&=\dfrac{\mu^2}{4\omega_0^2}\left[2\pi\mu\Omega_0-2\sqrt{\omega_0}x_0
\left(\dfrac{-2}{2\kappa+1}+\pi^2(2\kappa+1)\right)\right]^2,\label{regime1D=0B}\\
C&=\dfrac{4\mu^2}{\omega_0},\label{regime1D=0C}
\end{align}
where $x_0=(x_1-\mu\pi\sqrt{\omega_0}/2)$. On the other hand, the trapping energy of each particle in the potential well has the proper that $\bar{E}_k^\uparrow+\bar{E}_k^\downarrow\sim |\alpha_1|^2+|\alpha_2|^2=x_1^2+x_2^2
+y_1^2+y_2^2=2x_0^2+\mu^2\pi^2\omega_0/2+y_1^2+y_2^2
\geqslant2x_0^2+\mu^2\pi^2\omega_0/2+2\mu^2\Omega_0^2/\omega_0
\geqslant\mu^2\pi^2\omega_0/2+2\mu^2\Omega_0^2/\omega_0$,
with the second last equality sign holds only if $y_1=y_2=\mu\Omega_0/\sqrt{\omega_0}$. This property of the input state shows at least two facts: First, the trapping energy must large enough, or the measurement procedure in this part cannot be proceeded. Second, the sign of $x_0$ makes no difference when calculating the trapping energy. So when the trapping energy is fixed, we can always choose a negative $x_0$ in Eq.~(\ref{regime1D=0B}) and hence $B$ would have a positive correlation with $|x_0|$. As $y_1, y_2$ don't affect the measurement results, for a fixed trapping energy input state, we'd better let $y_1=y_2=\mu\Omega_0/\sqrt{\omega_0}$ that makes $|x_0|$ and as well $B$ largest. On the other hand, as the value of $C$ in Eq.~(\ref{regime1D=0C}) has no concern with the input state, hence the optimal uncertain of $\Omega$ would always reach the same SQL in this part no matter what coherent input state we choose, which also embodies a sense of stability. The best measurement result of $\omega(\Omega)$ is expressed as
\begin{align}
(\delta^2\omega_{\rm r})_{\rm m}\!&\sim\!\frac{1}{N^2}\cdot\frac{1}
{4\mu^2\left[(-2/\kappa_0\!+\!\pi^2\kappa_0)\sqrt{\omega_0}r_0
\!+\!\mu\pi\Omega_0\right]^2},\label{regime1D=0omega}\\
(\delta^2\Omega_{\rm r})_{\rm m}&\sim\frac{1}{N}\cdot\frac{\omega_0}{16\mu^2\Omega_0^2}.\label{regime1D=0Omega}
\end{align}
where $r_0=\sqrt{(r^2-\mu^2\pi^2\omega_0/2-2\mu^2\Omega_0^2/\omega_0)/2}$ is the maximum $|x_0|$ with $r^2=|\alpha_1|^2+|\alpha_2|^2$ indicates the trapping energy and $\kappa_0=2\kappa+1$ refers to any positive odd number. The ultimate uncertainty of $\omega$ is scaled by the HL now and it's clearly larger $r$ or higher trapping energy and larger $\kappa_0$ or longer evolution time would lead to more precise result. If the trapping energy of the input particles is given, namely, $r$ is fixed, to obtain the minimum value of Eq.~(\ref{regime1D=0omega}), the true values of $\omega_0, \Omega_0$ that are selected to measure around should be
\begin{eqnarray}
\label{omegaOmega}
\omega_0\sim\dfrac{r^2}{\pi^2\mu^2},~~\Omega_0\sim\dfrac{\kappa_0r^2}{2\mu^2
\sqrt{\pi^4\kappa_0^4-3\pi^2\kappa_0^2+4}}.
\end{eqnarray}
And the minimum uncertain of $\omega$ would be read as
\begin{eqnarray}
\label{omegalast}
(\delta^2\omega_{\rm r})_{\rm M}\sim\dfrac{1}{N^2}\cdot\dfrac{\pi^2\kappa_0^2}{r^4(\pi^4\kappa_0^4-3\pi^2\kappa_0^2+4)}.
\end{eqnarray}\\
\par
\noindent In summary, in the Condition ${\rm \RNum{2}}$ part, we can choice either of the two measured parameters be scaled by the HL and the measurement precision can surpass the one that obtained in Condition ${\rm\RNum{1}}$.
\par

\section{Conclusion}
\label{conclusion}
Our work in this paper reveals some inherent features of the Sagnac interferometer in the multiparameter estimation process. The general form of the input state is a quantum superposition of all particles that trapped in a harmonic potential spin up with all of them spin down. The measured parameters we choose is the rotation frequency $\Omega$ of the sagnac system and the trapping frequency $\omega$. The general expressions of the measurement precision are provided. We find no matter what the motional state of the particle is in the trap, the two measurement precisions cannot reach the HL simultaneously, only one can get the HL while the other can only be scaled by the SQL. By considering the constraint conditions that saturate the quantum Cram$\acute{e}$r-Rao bound, we can choose which measurement result approach the HL. The measurement precision of $\Omega$ here is better than the one in single parameter scheme, which may be useful for actual experiment researchers. The influence of the input to the measurement precision is embodied in the trapping energy, higher trapping energy would lead to more precise measurement result.

\vspace*{-1.ex}
\acknowledgments
\vspace*{-1.5ex}
This work was supported by the National Key Research and Development Program of China (Grants No. 2017YFA0304202 and No. 2017YFA0205700), the NSFC (Grants No. 11875231 and No. 11935012), and the Fundamental Research Funds for the Central Universities through Grant No. 2018FZA3005.





%

\end{document}